\documentclass[12pt]{iopart}

\usepackage{cite}
\usepackage{graphicx}
\usepackage{bm}
\usepackage{hyperref}

\usepackage[usenames,dvipsnames,x11names]{xcolor}

\newcommand{\be}{\begin{equation}}
\newcommand{\ee}{\end{equation}}
\newcommand{\bea}{\begin{eqnarray}}
\newcommand{\eea}{\end{eqnarray}}
\newcommand{\ket}[1]{\left\vert #1    \right\rangle }
\newcommand{\bra}[1]{\left\langle   #1  \right\vert}

\newcommand{\W}{\Omega}

\begin{document}
\title{Production of Fock Mixtures in Trapped Ions for Motional Metrology}
\author{Antonis Delakouras$^1$, Daniel Rodr\'iguez$^{2,3}$ and Javier Cerrillo$^4$\footnote{Corresponding author: javier.cerrillo@upct.es}}
\address{$^1$Humboldt Universit\"at zu Berlin, Berlin, Germany}
\address{$^2$Departamento de F\'isica At\'omica, Molecular y Nuclear, Universidad de Granada, 18071, Granada, Spain}
\address{$^3$Centro de Investigaci\'on en Tecnolog\'ias de la Informaci\'on y las Comunicaciones, Universidad de Granada, 18071 Granada, Spain}
\address{$^4$\'Area de F\'isica Aplicada, Universidad Polit\'ecnica de Cartagena, Cartagena 30202 Spain}
\begin{abstract}
We present a protocol to produce a class of non-thermal Fock state
mixtures in trapped ions. This class of states features a clear metrological
advantage with respect to the ground state, thus overcoming the standard
quantum limit without the need for full sideband cooling and Fock-state preparation on a narrow electronic
transition. The protocol consists in the cyclic repetition of red-sideband, measurement and preparation laser pulses.
By means of the Kraus map representation of the protocol, it is possible
to relate the length of the red sideband pulses to the specific class
of states that can be generated. With the help of numerical simulations,
we analyze the parametric regime where these states can be reliably
reproduced.
\end{abstract}

\section{Introduction}

Trapped ions are a platform of reference for the implementation and
testing of quantum information protocols \cite{Cirac1995,Wineland2012}, with several recent achievements in the quantum computing race \cite{Friis2018}. Beyond quantum computation,
quantum logic spectroscopy \cite{Schmidt2005} has opened up a useful avenue in quantum metrology \cite{Wineland2011}, both in the context of optical clocks \cite{Ludlow2015} and force sensing \cite{Biercuk2010,Ivanov2016,Shaniv2017}. Nevertheless, non-vanishing fluctuations of the motional ground state set a fundamental
limit (standard quantum limit, SQL) in the precision
of many quantum sensors. 

An ongoing effort to improve the sensing capabilities of trapped ions is
underway, with promising strategies arising in recent times. One of these consists in the use of non-thermal motional states. Squeezed states were one of the first workarounds to SQL that were proposed and implemented to great success \cite{LIGO2013} in the photonic context, and has been proposed \cite{Heinzen1990, Cirac1993} and implemented \cite{Meekhof1996, Home2016, Burd2019} in the context of trapped ions. In order to avoid the accurate control of the phase of squeezed states with respect to the measured force, excited Fock states ($n=1,2,...$)
have recently been proposed \cite{Wolf2019} as a means to achieve metrological
advantages with respect to the ground state. Their production
involves ground state cooling \cite{Wineland1975}, often via a transition that resolves
motional Fock states and therefore severely delays the cooling process.
Although higher cooling rates can be obtained by applying electromagnetically induced transparency \cite{Morigi2000}, they
cannot be arbitrarily increased \cite{Scharnhorst2018,Cerrillo2018} unless more elaborate implementations are considered \cite{Cerrillo2010}.

\begin{figure}
\centering\includegraphics[width=.6\columnwidth]{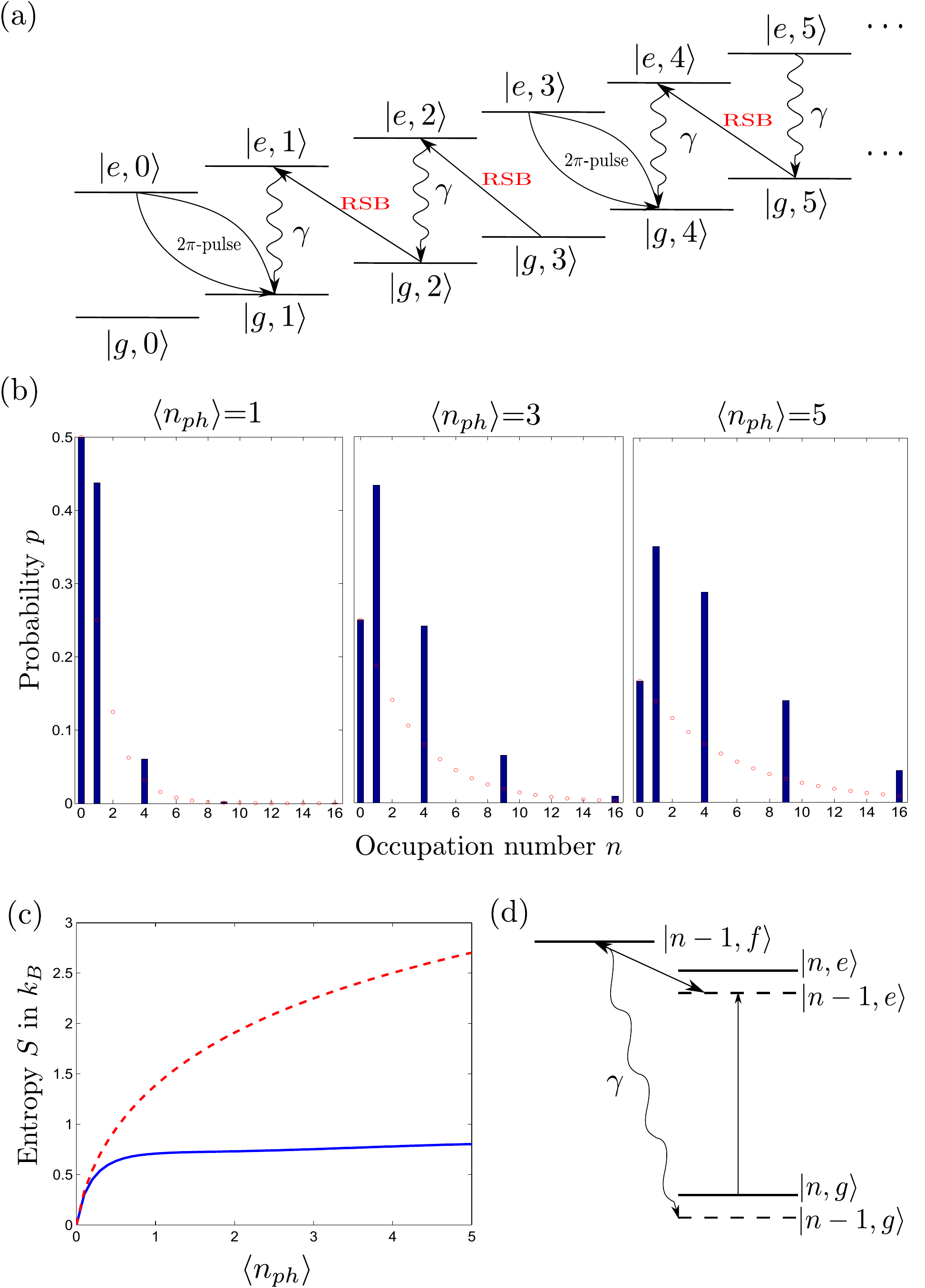}
\caption{(a) Production process and energy levels of a trapped state
for $n_{0}=1$. (b) Trapped (blue bars) and thermal (red dots) probability distributions
for various values of $\left\langle n_{ph}\right\rangle $ and $n_{0}=1$
(associated with the trap series $0,1,4,9,16,\dots$). (c) Thermal and
trapped state entropy for $n_{0}=1$ as a function of $\left\langle n_{ph}\right\rangle $. (d)  Experimental realisation through a third energy
level with a fast decay rate back to the electronic ground state.
\label{fig:n01}}
\end{figure}

Here we analyze a novel and simple protocol for the production of Fock
state mixtures that circumvents the need for ground-state laser cooling and repeated blue-sideband cycles to prepare a specific Fock state.
These mixtures, that we term \emph{trapped
states}, retain a significant metrological advantage with respect to the ground state. They feature a reduced entropy with respect to the states they were created from and, in addition, their parity can be well defined.
In this paper we explore in detail the production process as well
as the properties of trapped states, which can be generated with red-sideband excitations although also blue-sideband excitations can be used.

The paper is structured as follows. In sec.~\ref{sec:Model} we present the protocol formally, we define the concept of trapped states,
and also analyze their form in the special case of an initially thermal
ion. Sec.~\ref{sec:Quantum-Measurements-and} introduces the Kraus
map representation of the protocol. In sec.~\ref{sec:Displacement-Sensitivity-and}
we explore the applicability of the trapped state in the context of
quantum metrology and displacement sensitivity. In sec.~\ref{sec:Blue} we extend the idea to the use of blue-sideband excitations. Finally, in sec.~\ref{sec:Numerical-Simulation-of}
we simulate the results numerically under realistic conditions by
employing a Lindblad master equation and provide the final conclusions of the work.

\section{Implementation of the Protocol\label{sec:Model}}

We consider implementation of our protocol in a trapped-ion architecture.
Let us model the electronic degrees of freedom of a trapped ion with
a two-level system: an electronic ground state $\left|g\right\rangle $
and an excited state $\left|e\right\rangle $, with transition frequency
$\omega$. Its motional degrees of freedom are approximated by a quantum
harmonic oscillator of natural frequency $\nu$. Control is exerted by means
of a laser field of frequency $\omega_{l}$ that induces Rabi oscillations
of frequency $\Omega$ and is characterized by a Lamb-Dicke parameter
$\eta$ \cite{Neuhauser1978}. The Hamiltonian in the rotating wave approximation (RWA)
with respect to $\omega_{l}\sigma_{z}/2$ and taking $\hbar=1$ is
\begin{equation}
H=\frac{\Delta}{2}\sigma_{z}+\nu a^{\dagger}a+\frac{\Omega}{2}\left[\sigma^{+}D\left(i\eta\right)+\sigma^{-}D^{\dagger}\left(i\eta\right)\right],\label{eq:Hf}
\end{equation}
where $\Delta=\omega-\omega_{l}$ is the detuning of the laser with
respect to the transition, $a$ is the annihilation operator of the
harmonic oscillator, $D\left(\alpha\right)=\exp\left(\alpha a+\alpha^{*}a^{\dagger}\right)$
is the displacement operator, $\sigma^{+}=\left|e\right\rangle \left\langle g\right|$,
$\sigma^{-}=\left(\sigma^{+}\right)^{\dagger}$ are the spin raising
and lowering operators respectively and $\sigma_{z}=1-2\sigma^{-}\sigma^{+}$.
The Lamb-Dicke regime is quantitatively expressed by $\eta\sqrt{\left\langle n_{ph}\right\rangle}\ll 1$, where $\left\langle n_{ph}\right\rangle$ is the average phonon number. In this regime,
transitions that modify the motional state by more than a single phonon
are strongly suppressed. In this limit, setting the detuning $\Delta=\nu$,
and additionally performing a RWA with respect to $\nu\left(a^{\dagger}a+\sigma_{z}/2\right)$,
yields the red sideband (RSB) Hamiltonian
\begin{equation}
H_{RSB}=\frac{i\eta\Omega}{2}\left(a\sigma^{+}-a^{\dagger}\sigma^{-}\right),\label{eq:RSB Hamiltonian}
\end{equation}
also known as the the Jaynes-Cummings Hamiltonian \cite{Jaynes1963}.
It generates Rabi oscillations between states $\left|g,n\right\rangle \leftrightarrow\left|e,n-1\right\rangle $
at a frequency $\eta\Omega\sqrt{n}$.

The protocol, that we call selective population trapping (SPT) protocol,
simply consists in the periodic alternation of RSB
laser pulses with measurement and preparation steps.
In particular, we consider three steps
\begin{enumerate}
\item A RSB pulse is applied for a time $\tau=2\pi/(\eta\Omega\sqrt{n_{0}})$,
where $n_{0}>0$ is an integer of our choice that determines the
form of the trapped state.
\item An unread measurement of the electronic state of the ion is performed
at time $\tau$.
\item The ion is projected back into its electronic ground state $\ket{g}$.
\end{enumerate}
The effect of this process is illustrated in fig.~\ref{fig:n01}(a).
Population of most Fock states cascades down analogously to a sideband-cooling
scheme. Nevertheless, the length of the sideband pulse $\tau$ matches
the red-sideband period of motional Fock states $\left|n_{0}m^{2}\right\rangle $,
where $m$ is any natural number. All states for which the sideband
pulse represents a full Rabi oscillation will remain trapped. Eventually,
motional Fock mixtures are generated that have the form
\begin{equation}
\mu_{tr}=\sum_{m=0}^{\infty}p_{tr}(m)\left|n_0 m^{2}\right\rangle \left\langle n_0 m^{2}\right|,\label{eq:trapped}
\end{equation}
where the initial population of all states below trap $\left|n_0(m+1)^{2}\right\rangle $
has been deposited in trap $\left|n_0 m^{2}\right\rangle $, so
that, for an initial motional state $\mu_{0}$, we have 
\be
p_{tr}(m)=\sum_{k=n_0 m^{2}}^{n_0(m+1)^{2}-1}\left\langle k\right|\mu_{0}\left|k\right\rangle .
\ee
For the particular case of an initially thermal distribution $\mu^{th}_0$ of inverse temperature $\beta$, we have $\left\langle k\right|\mu^{th}_0\left|k\right\rangle =(1-e^{-\beta\nu})e^{-\beta\nu k}$,
and the final trapped state distribution becomes $p_{tr}^{th}(m)=e^{-\beta\nu n_0 m^2}-e^{-\beta\nu n_0 (m+1)^{2}}$.
Some examples are presented in fig.~\ref{fig:n01}(b).

The state described by eq.~(\ref{eq:trapped}) is a non-thermal probability
distribution. The function $p_{tr}(m)$ may even be non-monotonous: the position of its maximum depends only on the initial state $\mu_{0}$ and the time
$\tau$, as it can be clearly seen in fig.~\ref{fig:n01}(b). Since
the protocol concentrates population in a few trapping levels, it
is expected to reduce the entropy of the state, as shown in fig.~\ref{fig:n01}(c).
A proof that this is always the case for an initially diagonal state
in the Fock basis can be found in the Appendix.

Periodic electronic state measurement and preparation (steps 2 and
3 of the protocol) may be implemented by using electronic shelving
techniques \cite{Stevens1998}: an additional laser resonantly couples
$\left|e\right\rangle $ to a higher excited level of the ion $\left|f\right\rangle $,
which has a fast decay rate back to the electronic ground state of
the system, see fig.~\ref{fig:n01}(d). This technique is also commonplace
in implementations of standard sideband cooling in order to increase
cooling rates.

\section{Kraus Maps Analysis and Steady State\label{sec:Quantum-Measurements-and}}

As a way to analyze the dynamics of the motional degrees of freedom,
we employ the Kraus sum representation of quantum processes \cite{Kraus1983}.
The effect of the sideband, measurement and preparation pulses are
summarized by Kraus maps $K_{e}$ or $K_{g}$, depending on the outcome
of the electronic state measurement. Disregarding the measurement
outcome, the unconditional evolution of the density matrix $\mu(\tau)$
of the motional degrees of freedom is 
\be
\mu(\tau)=\sum_{i=e,g} K_{i}\mu_{0}K_{i}^{\dagger}.
\ee
Kraus maps satisfy the condition $\sum_{i=e,g}K_{i}^{\dagger}K_{i}=\mathbf{{1}}.$
Under the described protocol, they are computed as $K_{i}=\left\langle i\right|U\left(\tau\right)\left|g\right\rangle $,
with $U\left(\tau\right)=e^{-iH_{RSB}\cdot\tau}$ being the unitary
evolution operator associated with the red sideband pulse. The final
expressions read

\begin{eqnarray}
K_{g} & =&\sum_{n=0}^{\infty}\cos\left(\sqrt{n}\frac{\eta\Omega}{2}\tau\right)\left|n\right\rangle \left\langle n\right|,\label{eq:K_g}\\
K_{e} & = &-\sum_{n=0}^{\infty}\sin\left(\sqrt{n+1}\frac{\eta\Omega}{2}\tau\right)\left|n\right\rangle \left\langle n+1\right|.\label{eq:K_e}
\end{eqnarray}
Since by step 3 of the protocol the electronic state is prepared back
into $\left|g\right\rangle $, the same set of Kraus maps can be used
to describe repeated iterations of the SPT-Protocol. This is a useful
property in order to extract the steady state of the Fock state populations.

From the structure of the Kraus maps it can be seen that the evolution of populations and coherences is decoupled. In particular, we may describe the stroboscopic evolution of the vector of populations $\mathbf{p}(t)=\left(p_{0},p_{1},...,p_{n},p_{n+1},...\right)^{T}$ [with $p_n=\bra{n}\mu(t)\ket{n}$ and $t=k\tau$ any integer multiple of $\tau$] by means of the dynamical map $\mathcal{E}\left(\tau\right)$.
It is a matrix whose components are related to the Kraus maps
\begin{equation}
\mathcal{E}_{mn}\left(\tau\right)=\sum_{i=e,g}\left\langle m\right|K_{i}\left(\tau\right)\left|n\right\rangle \left\langle n\right|K_{i}^{\dagger}\left(\tau\right)\left|m\right\rangle .\label{eq:dynam map def}
\end{equation}
The steady state populations $\mathbf{p}^{ss}$ satisfy the equation
$\mathbf{p}^{ss}=\mathcal{E}(\tau)\mathbf{p}^{ss}$, which implies
\begin{equation}
\sin^{2}\left(\sqrt{n}\frac{\eta\Omega}{2}\tau\right)p_{n}^{ss}=\sin^{2}\left(\sqrt{n+1}\frac{\eta\Omega}{2}\tau\right)p_{n+1}^{ss}.\label{eq:eq prob trap}
\end{equation}
Beyond the
trivial solution (which corresponds to $p_{0}^{ss}=1$ and $p_{n}^{ss}=0$ for any $n>0$), this equation illustrates the reason for the choice $\tau=2\pi/(\eta\Omega\sqrt{n_{0}})$,
since it is only for this case that additional solutions exist, corresponding to the trapped states.

\section{Displacement Sensitivity and Quantum Metrology \label{sec:Displacement-Sensitivity-and}}

We now analyze the metrological advantage trapped states can have
with respect to the ground state in the field of displacement sensitivity.
In the spirit of \cite{Wolf2019}, a phase-space displacement $\alpha$
is implemented by letting the ion interact with an external electric
field. This corresponds to the state transformation $\mu_{tr}\rightarrow D(\alpha)\mu_{tr}D^{\dagger}(\alpha)$
with $D(\alpha)=\exp(\alpha a^{\dagger}-\alpha^{*}a)$. In the simplest
approach, the interaction is interrupted by a state read-out measurement
for the motional state $\left|n\right\rangle $ of the ion. This measurement
carries some information about $\alpha$. In particular, with the
help of the overlap function $\xi\left(\alpha\right)=\textrm{tr}\left\{ \left|n\right\rangle \left\langle n\right|D(\alpha)\mu_{tr} D^{\dagger}(\alpha)\right\} $
between the initial and the displaced state one can express the Fisher
information of the measurement with 
\be
\mathcal{F}\left(\alpha\right)=\frac{1}{\xi\left(\alpha\right)\left[1-\xi\left(\alpha\right)\right]}\left[\frac{d\xi\left(\alpha\right)}{d\alpha}\right]^{2},
\ee
which can then be used to quantify the metrological gain in comparison
to the SQL as 
\be
g=\frac{\mathcal{F}_{Q}\left(\alpha\right)}{\mathcal{F}_{SQL}},
\ee
where $\mathcal{F}_{Q}\left(\alpha\right)=\max_\alpha\mathcal{F}\left(\alpha\right)$
stands for the quantum Fisher information and $\mathcal{F}_{SQL}$
is the Fisher information produced by the motional ground state $\left|0\right\rangle $.
The quantum Fisher information is then directly linked to the achievable
measurement sensitivity $\Delta\alpha$ by means of the Cramer-Rao
bound $\Delta\alpha^{CR}$, given by
\be
\Delta\alpha\geq\Delta\alpha^{CR}=\frac{1}{\sqrt{N\mathcal{F}_{Q}\left(\alpha\right)}}.
\ee
More details on the calculations of state overlap and the associated
Fisher information are included in the Appendix.

\begin{figure}
\centering
\includegraphics[width=.5\columnwidth]{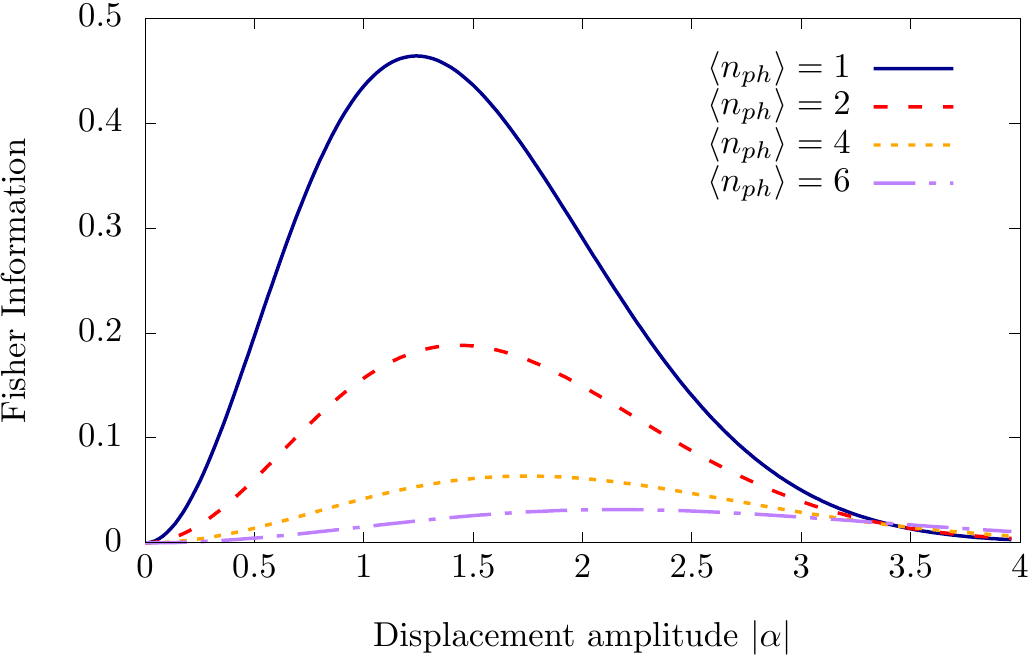}
\caption{Fisher information of the displacement amplitude $\left|\alpha\right|$ carried by a projective measurement of the motional ground state $\ket{0}$ given an initial thermal state of different values of $\left\langle n_{ph}\right\rangle $.\label{fig:ThermalFisher}}
\end{figure}

\subsubsection*{Thermal State}

The Fisher information of a thermal state characterized by an average
population $\left\langle n_{ph}\right\rangle $ is significantly lower
than the standard quantum limit and is therefore not appropriate for
metrological purposes. As illustrated in fig.~\ref{fig:ThermalFisher},
the maximum Fisher information decreases with an increasing $\left\langle n_{ph}\right\rangle $
and appears at larger values of the displacement amplitude
$\left|\alpha\right|$. This justifies the use of sideband cooling
to achieve higher sensitivities, although trapped
states can overcome the SQL without it, as we will now show.

\begin{figure}[t]
\noindent \begin{raggedright}
\includegraphics[width=1\textwidth]{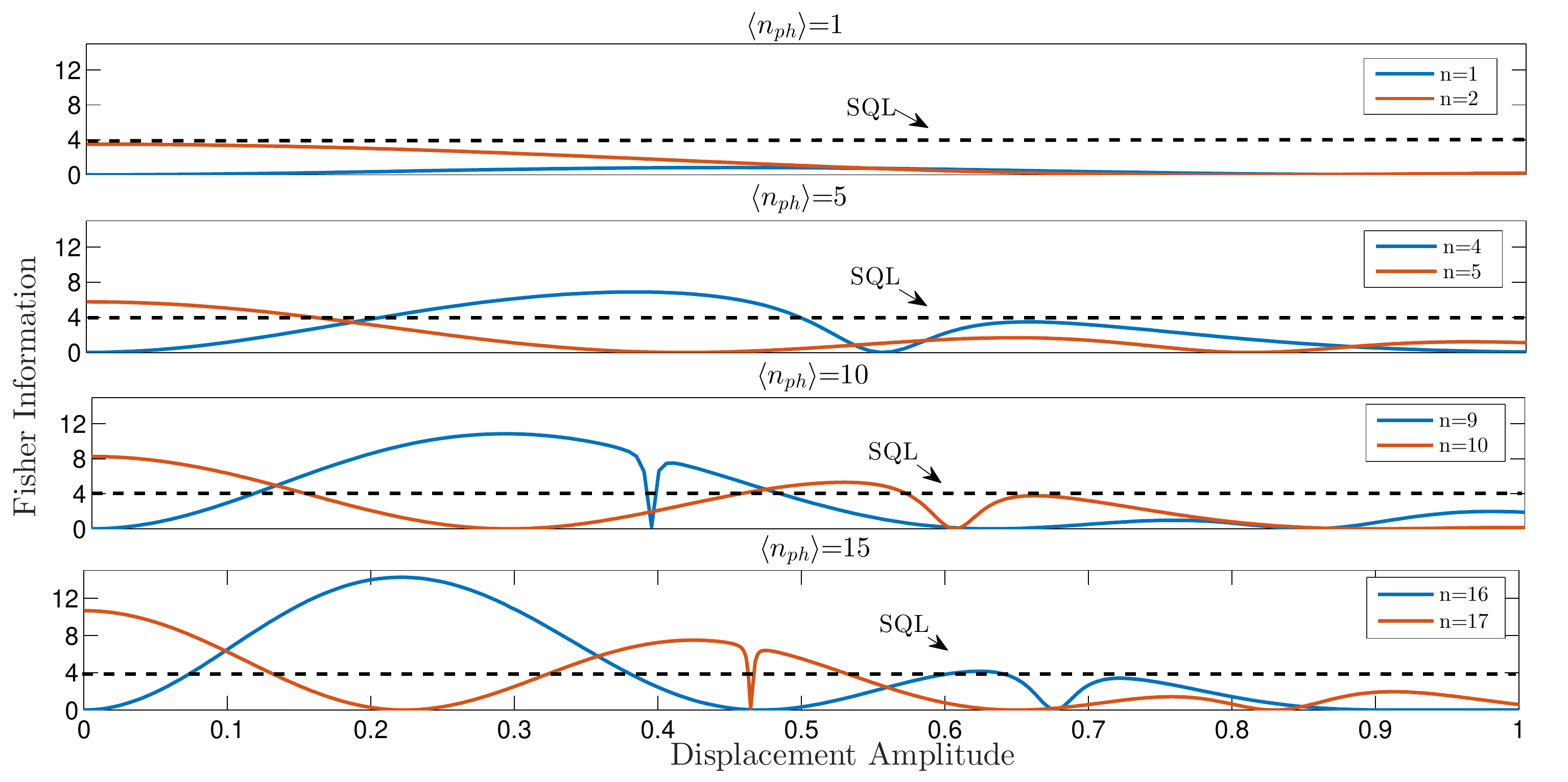}
\par\end{raggedright}
\caption{{\footnotesize{}\label{fig:Trapped-state-Fisher}}Trapped state Fisher
information as a function of the displacement amplitude $\left|\alpha\right|$
for different values of the average phonon number of the initial thermal
state $\left\langle n_{ph}\right\rangle $ and measured Fock state
$n$.}
\end{figure}

\subsubsection*{Trapped State}
Figure \ref{fig:Trapped-state-Fisher} shows the Fisher information
of a trapped state as a function of the displacement amplitude $\left|\alpha\right|$.
We consider trapped states created from thermal states with different
values of temperature represented by their initial average phonon
number $\left\langle n_{ph}\right\rangle $. For each value, two curves
are presented, each one corresponding to a measurement for a different
Fock state. The blue curve indicates a measurement of the most likely Fock state of the given trapped mixture $\mu_{tr}$, i.e.~the Fock state $\ket{n_0 m^2}$ with highest $p_{tr}(m)$. The red curve corresponds instead to a measurement of the next Fock state $\ket{n_0 m^2+1}$, which is not a member of the trapping series.
Red curves do not feature higher peaks in the Fisher information,
they do however result in higher Fisher information values for very
small displacement amplitudes. By comparison with fig.~\ref{fig:ThermalFisher},
it becomes apparent that trapped states produce significantly higher
Fisher information than thermal states. Additionally, the Fisher information
exceeds the SQL under certain conditions, as it can be seen in fig.~\ref{fig:Trapped-state-Fisher} for $\left\langle n_{ph}\right\rangle =5$,
$\left\langle n_{ph}\right\rangle =10$ and $\left\langle n_{ph}\right\rangle =15$.
As opposed to thermal states, achievable values of Fisher information
by trapped states increase with larger initial temperatures. This
proves that trapped states can be useful for metrological purposes,
since they can provide higher displacement sensitivities than the
motional ground state. 

In particular, we can expect the following metrological gains: 
\begin{eqnarray*}
\left\langle n_{ph}\right\rangle  & =5:\textrm{ }\textrm{ }g_{_{SQL}}=\frac{\mathcal{F}_{tr}\left(\alpha\approx0.4\right)}{\mathcal{F}_{SQL}}\approx1.75\mathrm{~dB},\\
\left\langle n_{ph}\right\rangle  & =10:\textrm{ }g_{_{SQL}}=\frac{\mathcal{F}_{tr}\left(\alpha\approx0.3\right)}{\mathcal{F}_{SQL}}\approx2.75\mathrm{~dB},\textrm{ and}\\
\left\langle n_{ph}\right\rangle  & =15:\textrm{ }g_{_{SQL}}=\frac{\mathcal{F}_{tr}\left(\alpha\approx0.24\right)}{\mathcal{F}_{SQL}}\approx3.5\mathrm{~dB}.
\end{eqnarray*}
It is worth stressing that trapped states have finite entropy [see fig.~\ref{fig:n01}(c)] and still are expected to provide a metrological gain with respect to the pure ground state (of zero entropy).

\section{Blue-Sideband Trapping\label{sec:Blue}}

Here we extend the idea of selective population trapping for the case
of a blue-sideband (BSB) excitation. The SPT protocol stays as defined
in sec.~\ref{sec:Model}, with the only exception being that now
a BSB transition is implemented in the first step. Since all the analytical
derivations for this section are formally identical to the ones detailed
in the RSB case, we are going to simply present the results in a compact
form and only focus on the differences between the two processes.
The Hamiltonian in this case is also known as anti-Jaynes-Cummings
Hamiltonian and is given in the RWA by 
\be
H_{BSB}=-\frac{i\eta\Omega}{2}\left(a^{\dagger}\sigma^{+}-a\sigma^{-}\right).
\label{aJC}
\ee
The corresponding Kraus maps have the form
\begin{eqnarray}
K_{g}^{'}&=\sum_{n=0}^{\infty}\cos\left(\sqrt{n+1}\frac{\eta\Omega}{2}\tau\right)\left|n\right\rangle \left\langle n\right|, \\
K_{e}^{'}&=\sum_{n=0}^{\infty}\sin\left(\sqrt{n+1}\frac{\eta\Omega}{2}\tau\right)\left|n+1\right\rangle \left\langle n\right|.
\end{eqnarray}
Solving for the steady state of these Kraus maps, the following equations arise
\be
\sin^{2}\left(\sqrt{n+1}\frac{\eta\Omega}{2}\tau\right)p_{n}^{ss}=\sin^{2}\left(\sqrt{n}\frac{\eta\Omega}{2}\tau\right)p_{n-1}^{ss}.
\ee
For a BSB-pulse time $\tau=2\pi/(\eta\Omega\sqrt{n_0})$, the final mixture now has the form
\begin{equation}
\mu_{tr}'=\sum_{m=1}^{\infty}p_{tr}'(m)\left|n_0 m^{2}-1\right\rangle \left\langle n_0 m^{2}-1\right|,\label{eq:trappedBSB}
\end{equation}
A difference in comparison to the RSB case is that now the first member of the
trapping series can be arbitrarily chosen and is not necessarily the
motional ground state $\ket{0}$.

A very interesting feature of the trapped state produced by a BSB
excitation is that one can reduce the entropy of the motional degree of freedom (the proof of this is
formally equivalent to the one presented in the appendix) by increasing
its energy. Additionally, with the BSB version of the
protocol it is possible to create mixtures of definite odd Fock state
parity, whereas RSB mixtures can be created with definite even parity.

\section{Numerical Simulation\label{sec:Numerical-Simulation-of} }

So far, the SPT protocol has been presented analytically, assuming
a Jaynes-Cummings Hamiltonian [eq.~(\ref{eq:RSB Hamiltonian})] or an anti-Jaynes-Cummings Hamiltonian [eq.~(\ref{aJC})].
Additionally, both the measurement and the spontaneous decay of the
ion back to the electronic ground state have been assumed to be instantaneous.
In this section, we test the predictions in a more realistic scenario
by performing numerical simulations that include both finite ground
state preparation time and the full Hamiltonian [eq.~(\ref{eq:Hf})].

\subsection*{Lindblad Master Equation and Evolution in Two Steps\label{subsec:Linblad-master-equation}}

In this numerical simulation we model the SPT-Protocol as a two-step
process: (1) the unitary evolution of the closed electronic-motional
system for time $\tau$ and (2) the unread measurement followed by
the spontaneous decay of the ion due to photon emission taking an
extra time $\tau_{e\rightarrow g}$. This is done by means of a master equation in Lindblad form \cite{Lindblad1976} as given in \cite{Haroche2006}. It acts on the \emph{total} density
matrix $\rho$, involving both electronic and motional degrees of freedom, and is expressed by the following equation
\begin{eqnarray}
\nonumber\frac{d}{dt}\rho&=-i\left[H,\rho\right]+\frac{\Gamma}{2}\left(2\sigma^{-}\rho\sigma^{+}-\sigma^{+}\sigma^{-}\rho-\rho\sigma^{+}\sigma^{-}\right)\\
&\equiv\mathcal{L}(\Omega,\Gamma)\rho,
\end{eqnarray}
where $\Gamma$ is the decay rate between the two qubit states and
$\mathcal{L}$ stands for the Liouvillian of the system. By using
the full Hamiltonian from eq.~(\ref{eq:Hf}), we include the effects of off-resonant
carrier and blue sideband excitations, as well as all higher-order
Lamb-Dicke terms.

\begin{figure}[h]
\noindent \centering{}\includegraphics[width=.5\columnwidth]{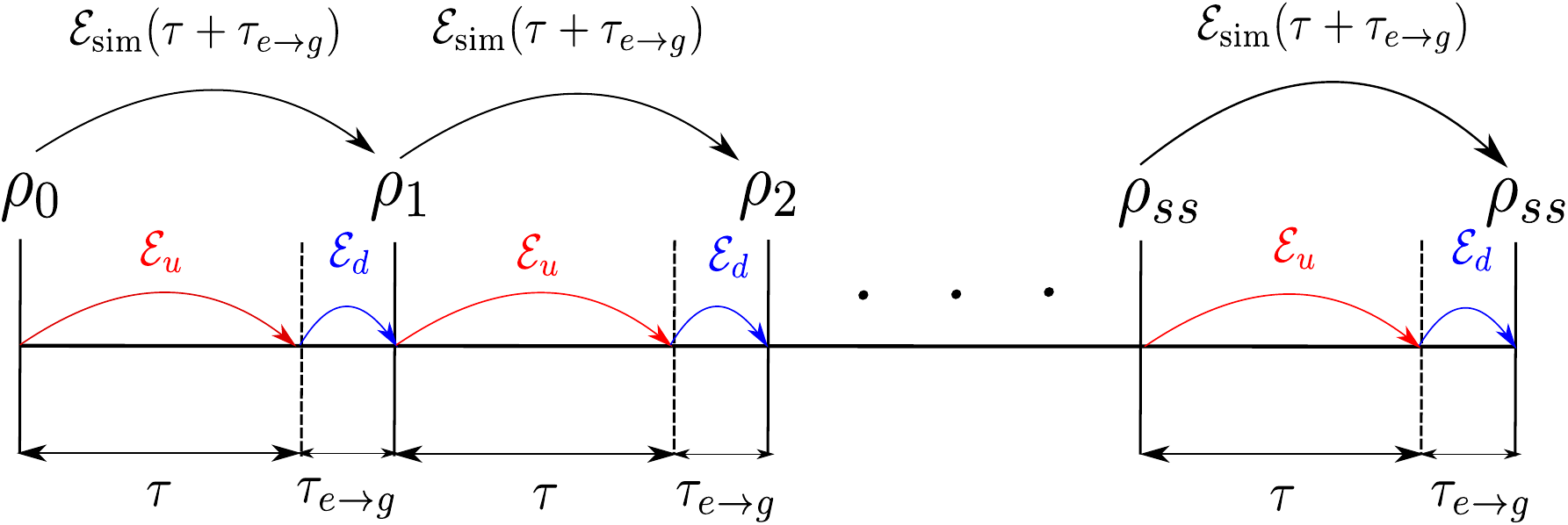}
\caption{Schematics for the two-step process describing one cycle
of the numerical simulation of the SPT-Protocol. The dashed vertical
lines represent the end of the unitary interaction and the solid lines
the density matrix after a whole cycle of the protocol.\label{fig:Numerics}}
\end{figure}

The dynamical map describing the evolution of $\rho$ is the result of the product $\mathcal{E}_{sim}\left(\tau+\tau_{e\rightarrow g}\right)$=$\mathcal{E}_{d}\left(\tau_{e\rightarrow g}\right)\cdot\mathcal{E}_{u}\left(\tau\right)$,
where $\mathcal{E}_{u}\left(t\right)=\exp[\mathcal{L}(\Omega,0)t]$
represents the unitary part of the evolution (with $\Gamma=0$) and
$\mathcal{E}_{d}(t)=\exp[\mathcal{L}(0,\Gamma)t]$ the dissipative
part of the evolution (with $\Omega=0$). This is illustrated in fig.~\ref{fig:Numerics}. 

\subsection*{\label{res and disc}Results and Discussion}

We compare the analytical prediction for the population vector $\mathbf{p}_{R}$
after $R$ repetitions of the SPT protocol as provided by the ideal
dynamical map $\mathcal{E}(\tau)$ from eq.~(\ref{eq:dynam map def}),
$\mathbf{p}_{R}=\mathcal{E}^{R}(\tau)\mathbf{p}_{0},$ and as computed
from the numerical simulation, $\rho\left(R\tau+R\tau_{e\rightarrow g}\right)=\mathcal{E}_{sim}^{R}(\tau+\tau_{e\rightarrow g})\rho\left(0\right).$
This facilitates identification and analysis of the three parametric
requirements ($R\gg1$ , $\eta\ll1$ and $\Omega\ll\nu$) necessary
for SPT to work. For simplicity, throughout this section we use {\small{}$n_{0}=1$}.
Simulations were performed with a cutoff $\textrm{dim}_{m}=14$ for
the motional Hilbert space (maximum Fock state number), which was
found to produce sufficient convergence.

\subsubsection*{Number of Repetitions}

First, varying the number of repetitions $R$ and comparing the results
leads to an understanding of how many applications of the SPT-Protocol
are necessary in order to observe the effect of population trapping
expected in the steady state. For the parametric regime in which trapped states are well approximated  ($\eta\ll1$ and $\Omega\ll\nu$), we find that $R=30$ is sufficient to reach a reasonable approximation to the steady state,
since higher values do not appreciably modify the distribution.



\begin{figure}
\centering
\includegraphics[width=.5\columnwidth]{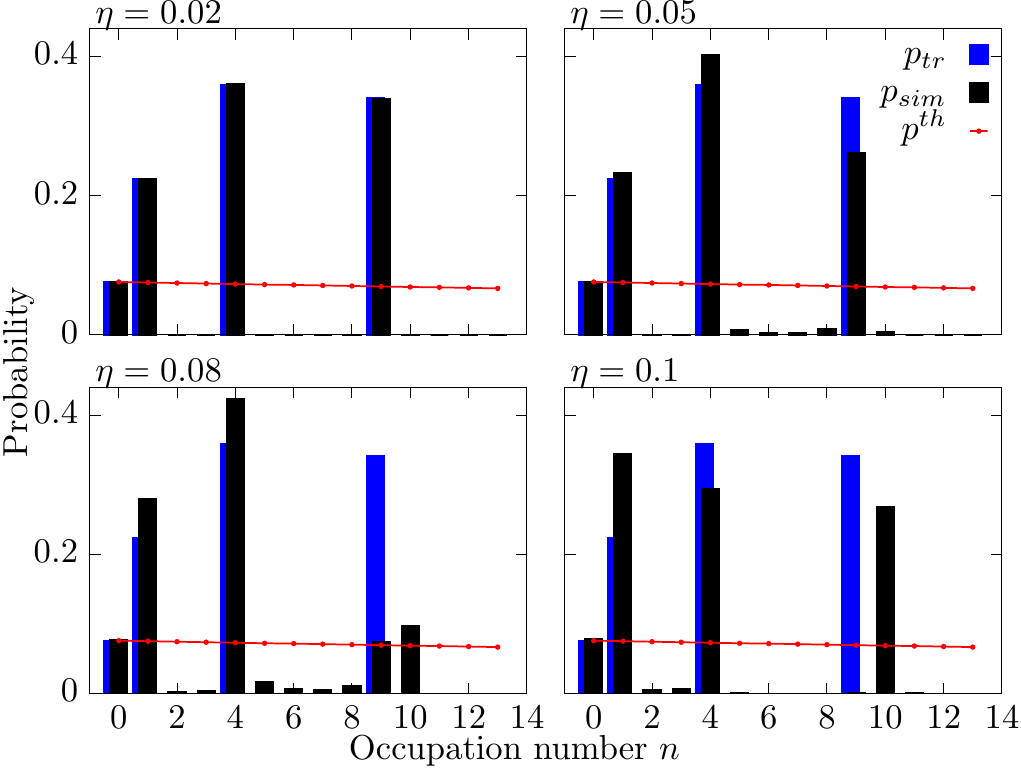}
\caption{\label{etavar}Comparison between the analytical predictions $p_{tr}$
(blue) and simulated results $p_{sim}$ (black) for the probability distribution
of Fock populations for $R=30$ repetitions of the protocol applied
on an initially thermal state $p^{th}$ (red) for various values of the Lamb-Dicke
parameter. The following parameter values have been used: $n_{0}=1$,
$\Omega=10^{-4}\nu$, $\beta=0.01\nu^{-1}$, $\delta=\nu$ and
$\Gamma_{e\rightarrow g}=1000\nu$.}
\end{figure}

\subsubsection*{Lamb-Dicke Parameter }

 A small $\eta$ leads to population distributions significantly closer to
the analytical predictions, as can be seen in fig.~\ref{etavar}. This is due to the Lamb-Dicke approximation $\eta\sqrt{\left\langle n_{ph}\right\rangle}\ll1$, which
loses its validity both as $n$ or $\eta$ increase. For
values of $\eta\leq0.02$, the simulated results completely match
the analytical ones after about 20 repetitions of the protocol. Values
in the range $0.02<\eta<0.06$ only approach the predictions within
a margin of about $10\%$, while for higher values the two predictions
are completely incompatible. For larger $\eta$, the Lamb-Dicke
approximation is no longer valid and one needs to account for higher
order terms of the form $a^{\dagger(m+n)}a^{m}$, which couple states
with quantum numbers that vary by $n$, i.e. $\left|g,m\right\rangle \rightarrow\left|e,m+n\right\rangle $.
The role of $n$ in the loss of validity of the Lamb-Dicke approximation is also clear from the simulation: as $\eta$ increases, lower values of $n$ are affected. As shown in the top right subplot of fig.~\ref{etavar}, a small increase in $\eta$ primarily
affects trapping state $n=9$, while further increase of $\eta$ (bottom right) unstabilizes
the $n=4$ trapping state as well. Meanwhile the $n=1$ trapping state remains stable
and accumulates the escaped population.

\begin{figure}
\centering
\includegraphics[width=.5\columnwidth]{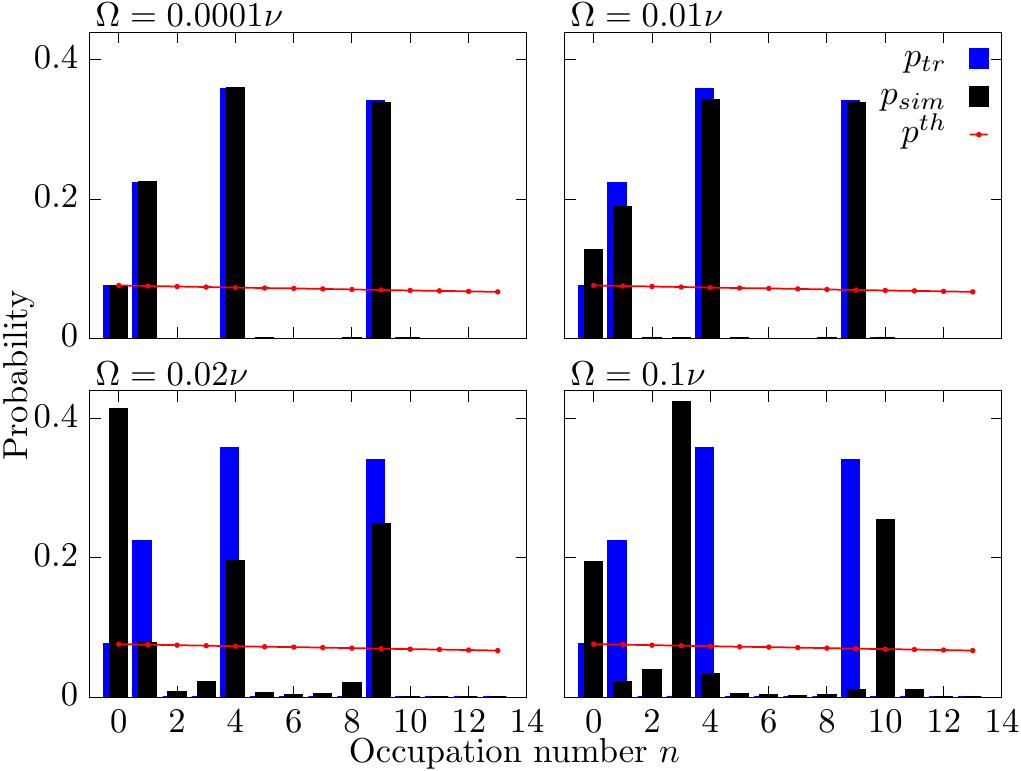}\caption{\label{omegavar}Comparison between the analytical predictions $p_{tr}$ (blue) and simulated results $p_{sim}$ (black) for the probability distribution
of Fock populations for $R=30$ repetitions of the protocol applied
on an initially thermal state $p^{th}$ (red) for various values of the Rabi
frequency. For all the subplots, the following parameter values have been
used: $n_{0}=1$, $\eta=0.02$, $\beta=0.01\nu^{-1}$, $\delta=\nu$
and $\Gamma_{e\rightarrow g}=1000\nu$.}
\end{figure}

An interesting effect that can be observed in the bottom right subplot
of fig.~\ref{etavar}, is the formation of a trap at $n=10$ for
$\eta=0.1$, which cannot be explained within the Jaynes-Cummings
type model employed in the previous sections. The trapping at that state is very persistent, and is present even after 1000 repetitions of the protocol. We suspect this behaviour to be due to the Debye-Waller reduction factor of the Lamb-Dicke parameter due to high-order effects \cite{Wineland1998} and will be investigated elsewhere. At any rate, it is an indication that the concept of selective population trapping could be extended to models that account for all Lamb-Dicke orders.

\subsubsection*{Rabi Frequency}

We now focus on how different Rabi frequencies influence the results,
as shown in fig.~\ref{omegavar}. Increased $\Omega$ leads to some
population escaping the traps due to \emph{off-resonant carrier excitation},
as the corresponding term in the Hamiltonian of eq.~(\ref{eq:Hf}) becomes more
important. In order to remain within the regime of validity of the RWA underlying $H_{RSB}$, we observe that Rabi and trap frequencies
require a separation of approximately three orders of magnitude. Values of $\W$ that
are only two orders of magnitude larger than $\nu$ quickly destabilize the traps.
In contrast to an increased Lamb-Dicke parameter, this affects the
lower lying traps first. It has also been observed that, for higher Rabi frequency
values that are within the range of about $\left(0.5-1\right)\%$
of the trap frequency, the analytical result is better reproduced for
about twenty repetitions of the protocol, instead of the thirty shown in the figures.

In conclusion, population trapping as described in sec.~\ref{sec:Model}
is experimentally applicable in the regimes given by $\eta\leq0.02$
and $\Omega\leq0.005\nu$, which directly results from the approximations
made in the derivation of the Hamiltonian in eq.~(\ref{eq:RSB Hamiltonian}).

\section*{Conclusions}

We have presented a protocol for the creation of a special class of
motional states in trapped ions. The protocol is simple and involves
the alternate concatenation of red or blue sideband pulses with measurement
and preparation pulses. The form of the generated state depends on
the duration of the sideband pulses and populates only Fock states
proportional to perfect square numbers. The presence in the mixture
of excited Fock states makes them especially suitable for motional
metrology, improving on the standard quantum limit without the need
for long sideband cooling pulses. The protocol works best with small values of the
Lamb-Dicke parameter in order to avoid deleterious high-order effects,
and small Rabi frequencies to suppress carrier effects. Such a protocol  might be applied in a Penning trap to measure the motional frequencies of a single ion and even extend it to the case of an unbalanced two-ion crystal where the motional modes of the crystal has to be probed \cite{Gutierrez2019,Cerrillo2021}.

\ack
J.C. acknowledges support from {\it Ministerio de Ciencia, Innovaci\'on y Universidades} (Spain) (``Beatriz Galindo" Fellowship BEAGAL18/00081). D.R. acknowledges support from Junta de Andaluc\'ia through the project P18-FR-3432.

\appendix

\section{Trapped and Thermal State Entropy}

The Von Neumann entropy \cite{Neumann1996} of the system described
by a density matrix $\mu$ is given by
\be
S=-k_{B}\textrm{tr}\left(\mu\ln\mu\right),
\ee
which for a diagonal density matrix simplifies to
\begin{equation}
S=-k_{B}\sum_{m=0}^{\infty}p_{m}\cdot\ln\left(p_{m}\right),\label{entropy}
\end{equation}
and essentially quantifies the degree of mixedness for a given state.
Since in a trapped state there are significantly fewer states populated,
one would expect a lower entropy compared to the thermal state that
was used to produce it. The following table shows the first few probabilities
involved in the sum of eq. (\ref{entropy}) for both a diagonal and
a trapped state, assuming $n_{0}=1$, in order to get an idea on how
they relate to each other.
\begin{center}
\begin{tabular}{|c|c|c|c|c|c|}
\hline 
Thermal & $p_{0}$ & $p_{1}$ & $p_{2}$ & $p_{3}$ & $p_{4}$\tabularnewline
\hline 
\hline 
Trapped & $p_{0}$ & $\sum_{m=1}^{3}p_{m}$ & $0$ & $0$ & $\sum_{m=4}^{8}p_{m}$\tabularnewline
\hline 
\end{tabular}\\
\par\end{center}
It is apparent that the contribution to entropy generated by three states with
probabilities $p_{1}$, $p_{2}$ and $p_{3}$ respectively is replaced in the trapped state case by the contribution of a single state of total probability $p_{1}+p_{2}+p_{3}$.

In general, we have trapped state probabilities 
\be
p_{tr}(m)=\sum_{k=A}^{B}p_{k},
\ee
where $A\equiv n_{0}\cdot m^{2}$ and $B\equiv n_{0}\left(m+1\right)^{2}-1$. 
The contribution to the entropy corresponding to a single
trap and the states until the next trap is
\begin{equation}
p_{tr}(m)\ln\left[p_{tr}(m)\right]=\ln\left[p_{tr}(m)\right]\sum_{k=A}^{B}p_{k}.\label{eq:entropy1}
\end{equation}
If we compare it to the contribution to the entropy of the same states
in the original distribution
\be
\sum_{k=A}^{B}p_{k}\ln\left(p_{k}\right)= p_{A}\ln\left(p_{A}\right)+ p_{A+1}\ln\left(p_{A+1}\right)+...+p_{B}\ln\left(p_{B}\right).\label{entropy2}
\ee
Considering that $p_{tr}(m)>p_{k}$ with $k$ between $A\equiv n_{0}\cdot m^{2}$
and $B\equiv n_{0}\left(m+1\right)^{2}-1$ leads to $\ln\left[p_{tr}(m)\right]>\ln\left(p_{k}\right)$.
In combination with equations (\ref{eq:entropy1}) and (\ref{entropy2})
we get
\be
p_{tr}(m)\ln\left[p_{tr}(m)\right]>\sum_{k=A}^{B}p_{k}\ln\left(p_{k}\right).
\ee
Summing over the remaining traps and multiplying with $-k_{B}$ yields the trapped entropy $S_{tr}$ and original entropy $S_0$ 
the predicted result 
\be
S_{tr}<S_{0}.
\ee

\section{State Overlap }

For a prepared state of the form $\mu_{0}=\sum_{m=0}^{\infty}p_{m}\left|m\right\rangle \left\langle m\right|$
the state overlap between the pure state and displaced prepared state
can be calculated by 
\begin{eqnarray*}
\xi\left(\alpha\right) & =\textrm{tr}\left\{ \left|n\right\rangle \left\langle n\right|D\left(\alpha\right)\sum_{m=0}^{\infty}p_{m}\left|m\right\rangle \left\langle m\right|D^{\dagger}\left(\alpha\right)\right\} \\
 & =\sum_{m=0}^{\infty} p_{m}\left|\left\langle n\right|D\left(\alpha\right)\left|m\right\rangle \right|^{2}.
\end{eqnarray*}
The general formula for the scalar product between two displaced Fock
states as given in \cite{Wuensche1991} is 
\be
\left\langle n\right|D^{\dagger}\left(\beta\right)D\left(\alpha\right)\left|m\right\rangle  
=\sqrt{\frac{m!}{n!}}\left(\alpha-\beta\right)^{n-m} \left\langle \beta\right|\left.\alpha\right\rangle \mathcal{L}_{m}^{n-m}\left\{ \left(\alpha-\beta\right)\left(\alpha^{*}-\beta^{*}\right)\right\} 
\ee
with $\left\langle \beta\right|\left.\alpha\right\rangle =\exp\left\{ \alpha\beta^{*}-\frac{1}{2}\left(\alpha\alpha^{*}+\beta\beta^{*}\right)\right\} $.

Inserting $\beta=0$ gives
\begin{equation}
\xi\left(\alpha\right)=e^{-\left|\alpha\right|^{2}}\sum_{m=0}^{\infty}p_{m}\frac{m!}{n!}\left|\alpha\right|^{2(n-m)}\left\{ \mathcal{L}_{m}^{n-m}\left(\left|\alpha\right|^{2}\right)\right\} ^{2},\label{eq:overlap}
\end{equation}
with $\mathcal{L}_{\mu}^{\nu}\left(x\right)$ being the generalized
Laguerre polynomials as defined in \cite{Haeringen1983}.

Inserting the thermal and trapped probability distribution in eq.~(\ref{eq:overlap}) gives the respective overlap expressions.

\section{Fisher Information}

The Fisher information is a measure of how quickly a probability distribution
$P\left(x|\theta\right)$ changes with respect to the parameter
$\theta$. In order to derive an analytical expression for the Fisher
information we follow the method from \cite{Wolf2019}. 

The precision of an estimation is limited by the Cramer-Rao
bound as 
\be
\Delta\theta_{est}\geq\Delta\theta_{CR}=\frac{1}{\sqrt{N\mathcal{F}\left(\theta\right)}},
\ee
where $\theta_{est}$ is an arbitrary estimator for $\theta$, $N$ is
the number of repeated measurements, and
\be
\mathcal{F}\left(\theta\right)=\sum_x\frac{1}{P\left(x|\theta\right)}\left[\frac{\partial P\left(x|\theta\right)}{\partial\theta}\right]^{2}
\ee
is the classical Fisher Information. In metrological applications,
one is interested in maximizing the precision estimation of the parameter
$\theta$. For this purpose, a minimum $\Delta\theta_{CR}$, or equivalently
a maximum $\mathcal{F}\left(\theta\right)$ is required.

The probability distribution $P\left(x|\theta\right)=\textrm{tr}\left\{ \Pi_{x}\mu\left(\theta\right)\right\} $
depends on the quantum state $\mu\left(\theta\right)$ and the choice
of the performed measurement, described by the projectors $\left\{ \Pi_{x}\right\} $.
For processes where the parameter $\theta$ is imprinted by a unitary
process, i.e. $\mu\left(\theta\right)=U\left(\theta\right)\mu_{0}U^{\dagger}\left(\theta\right)$,
the Fisher information has a lower bound given by
\begin{equation}
\mathcal{F}\left(\theta\right)\geq\frac{1}{\left(\Delta M\right)_{\mu\left(\theta\right)}^{2}}\left[\frac{\partial\left\langle M\right\rangle _{\mu\left(\theta\right)}}{\partial\theta}\right]^{2},\label{bound}
\end{equation}
where $\left\langle M\right\rangle _{\mu\left(\theta\right)}=\textrm{tr}\left\{ M\mu\left(\theta\right)\right\} $
is the mean value and $\left(\Delta M\right)_{\mu\left(\theta\right)}^{2}=\left\langle M^{2}\right\rangle _{\mu\left(\theta\right)}-\left\langle M\right\rangle _{\mu\left(\theta\right)}^{2}$
the variance of the measured observable $M=\sum_x x\Pi_{x}$. 

If there exist only two possible outcomes from a measurement, $x=0,1$,
this bound is tight. Considering that the probabilities must add up
to one, $P\left(0|\theta\right)=1-P\left(1|\theta\right)$ and the
variance becomes
\begin{equation}
\left(\Delta M\right)_{\mu\left(\theta\right)}^{2}=P\left(1|\theta\right)\left[1-P\left(1|\theta\right)\right].\label{eq:variance}
\end{equation}

Let us now consider the metrological protocol described in the manuscript. Starting with a state $\rho_{0}$, we are interested in how
a displacement $D(\alpha)$ affects the probability distribution of
the state. In other words, we investigate how sensitive the ion is
to that displacement. The higher the sensitivity (quantified by the
Fisher information), the more precise the parameter estimation is
for the parameter $\theta=\alpha$. The displacement operator is unitary
and transforms the density matrix as 
\be
\mu\left(\alpha\right)=D\left(\alpha\right)\mu_{0}D^{\dagger}\left(\alpha\right).
\ee
A projective measurement for a pure Fock state $\left|n\right\rangle \left\langle n\right|$
has only two possible outcomes, and therefore the bound (\ref{bound})
is tight and the measured observable $M$ takes the form 
\be
M=\sum_{x=0,1}x\Pi_{x}=\left|n\right\rangle \left\langle n\right|.
\ee
This results in a mean value of 
\be
\left\langle M\right\rangle _{\mu\left(\alpha\right)}=\textrm{tr}\left\{ M\mu\left(\alpha\right)\right\} =\xi\left(\alpha\right).
\ee
Combining this with $P\left(1|\alpha\right)=\textrm{tr}\left\{ \Pi_{1}\mu\left(\alpha\right)\right\} =\xi\left(\alpha\right)$
and the tightness of the bound (\ref{bound}) results in
\begin{equation}
\mathcal{F}\left(\alpha\right)=\frac{1}{\xi\left(\alpha\right)\left[1-\xi\left(\alpha\right)\right]}\left[\frac{d\xi\left(\alpha\right)}{d\alpha}\right]^{2}.
\end{equation}

\end{document}